\documentclass[12pt]{article}
\usepackage{amssymb,amsmath,subfigure,psfrag,epsfig,bm}
\usepackage[linesnumbered,ruled,vlined]{algorithm2e}
\title{Machine Learning for LiDAR-Based Navigation System}

\author{Farhad Aghili\thanks{email:faghili@encs.concordia.ca}}

\date{}
\begin{document}
\maketitle

\begin{abstract}
This paper presents a robust  6-DOF relative navigation by
combining the iterative closet point  (ICP)  registration
algorithm and a noise-adaptive Kalman filter (AKF) in a
closed-loop configuration together with  measurements from a
laser scanner and an inertial measurement unit (IMU). In this
approach, the fine-alignment phase of the registration is
integrated with the filter innovation step  for estimation
correction while the filter estimate propagation provides the
coarse alignment needed to find the corresponding points at the
beginning of ICP iteration cycle.  The convergence of the ICP
point matching is monitored by a fault-detection logic and the
covariance associated with  the ICP alignment error is estimated
by a recursive algorithm.  This ICP enhancement has proven to
improve  robustness and accuracy of the pose tracking performance
and to automatically recover correct alignment whenever the
tracking is lost. The Kalman filter estimator is designed so as to identify the
required parameters such as IMU biases and location of the
spacecraft center-of-mass (CoM).  
\end{abstract}

\section{Introduction}
Robust relative navigation systems are  critical for many current
and  near-future lunar or space exploration missions  to support
rendezvous, proximity operations and docking for both crewed and
un-crewed vehicles \cite{Aghili-2016c,Hoffmann-Gorinevsky-2007,Aghili-2022}. Fault-tolerant
is  a vital design requirement in development of  relative
navigation systems for these aerospace applications.  In particular, reliable relative pose information in full 6-dof is
required during approach and docking operations for visiting vehicles with the
International Space Station (ISS). It is deemed that the safety of
the controlled spacecraft during such proximity maneuvers
critically depends on the performance and robustness of the
relative navigation systems. Failure of  relative navigation
systems to provide continuous and accurate pose (position and
orientation) is considered as a critical hazard or even a
catastrophic hazard that can cause failure of the mission all
together.

There are different relative navigation sensors capable of
estimating the pose   of objects having relative motion.
Application of radar and altimetry for space-borne navigation
systems began more than half a century ago
\cite{Kriegsman-Bernard-1966}, while  X-ray pulsars for relative
navigation between two spacecraft in deep space is recently
presented \cite{Emadzadeh-Speyer-2011,Aghili-2019e,Liu-Fang-2015,Aghili-2012b}. Other
relative navigation methods focus on using Global Position System
(GPS) for determining both absolute and relative position and
attitude between two spacecraft
\cite{Aghili-Kuryllo-Okouneva-English-2010a,Corazzini-Robertson-1997,Wolfe-Speyer-2004,Aghili-2013,Almagbile-Wang-Ding-2010}.
There are also different vision systems capable of estimating the
pose  of two objects moving with respect to each other
\cite{Masutani-Iwatsu-Miyazaki-1994,
Samson-English-Deslauriers-Christie-2004,Mokuno-Kawano-2004,Aghili-Parsa-2008b,Aghili-2008c}.
Although  using radar or GPS for relative navigation systems are
with the advantage of long-range distance measurement, they are
with less resolution and precision compared with the vision-based
systems. Moreover, the advent of relatively low-cost and
commercially available laser range sensors and scanners, which has
been greatly exploited for autonomous navigation of robotic
vehicles
\cite{Shibata-Matsumoto-1996,Aghili-2011k,Lu-Tomizuka-2006,Leavitt-Sideris-2006,Guoqiang-Corradi-2011,Wang-Liu-2014,Simanek-Reinstein-2015,Aghili-2010f},
makes them preferred sensor of choice in relative navigation
systems. Moreover, it has been shown that laser scanners exhibit
acceptable robustness in the face of the harsh lighting conditions
of space. For insane, a rendezvous laser sensor  was used as the
primary navigation to perform unmanned autonomous rendezvous
docking experiments in the ETS-VII mission
\cite{Mokuno-Kawano-2004}. Vision algorithms for laser scanners
have been also developed for motion estimation of free-floating
objects to support a variety of on-orbit proximity
operations~\cite{Masutani-Iwatsu-Miyazaki-1994,English-Zhu-Smith-Ruel-Christie-2005,Aghili-Parsa-Martin-2008b,Aghili-Parsa-2007b,Aghili-Parsa-2008b,Aghili-Parsa-Martin-2008a,Aghili-Kuryllo-Okouneva-English-2010c,Aghili-Kuryllo-Okouneva-English-2010b}.
However, fault tolerance capability is not incorporated in these
vision algorithms to achieve robust pose-tracking performance.

The conventional vision-based pose estimation algorithms are
essentially 3D registration processes, by which the range data set
from different views are aligned in a common coordinate system
\cite{Kim-Hwang-2004}.  The iterative closest point (ICP) is  the
cornerstone of 3D vision-based pose estimation algorithm and one
of the most popular registration methods. The iterative procedure
minimizes a distance between point cloud in one dataset and the
closest points in the
other~\cite{Besl-Mckay-1992,Greenspan-Yurick-2003}. Typically, one
dataset is a set of 3D point-cloud acquired by scanning an object,
while the other one is a model set such as a CAD model of the same
object. The basic ICP algorithm has proven to be very useful in
the processing of range data \cite{Greenspan-Yurick-2003} and
subsequently a number of variations on the basic method have been
developed to optimize different phases of the algorithm
\cite{Chen-Medioni-1992,Rusinkiewicz-Levoy-2001,Godin-Laurendeau-Bergevin-2001,Greenspan-Yurick-2003}.
Recently, several improved algorithms of ICP by adopting Lie
groups have been reported in the literature
\cite{Du-Zheng-Ying-Liu-2010,Dong-Peng-Ying-Hu-2014}.
 The advantage of these algorithms is
that it converges monotonically to a local minimum from any given
initial parameters. Although the modified ICP algorithms improve the
convergence to local minima and sensitivity to outliers and
disturbances, they still suffer from slow convergence or even
divergence if proper initial pose estimate is not available
\cite{Li-Yin-Huang-2011}. Other  registration methods
include tangent-squared distance minimization (TDM)
\cite{Pottmann-Hung-Yang-2006}, squared distance minimization
(SDM) \cite{Wang-Pottmann-Liu-2006}, and adaptive distance
function (ADF) \cite{Li-Yin-Huang-2011}. Although  point-tangent
methods, i.e., TDM and SDM, reportedly exhibit faster convergent
rate than ICP does \cite{Pottmann-Hung-Yang-2006}, they have their
own drawbacks. For instance, the efficiency of SDM algorithm is
compromised by the heavy computation required to get the curvature
value of each discrete point, while TDM requires a good initial
estimate of initial pose otherwise registration may fail in
iteration \cite{Li-Yin-Huang-2011}. The shortest distance is more
generally defined in the ADF method than in the other traditional
methods. In fact, by varying a scalar coefficient, the ADF
distance function can be turned into the point-tangent distance
function of TDM or the point-point distance function of ICP
\cite{Li-Yin-Huang-2011}. This means that  TDM and ICP methods are
actually special cases of the ADF. Therefore, the ADF algorithm has the
potential to get a compromised performance with faster convergence
rate of ICP and improved robustness of TDM
\cite{Li-Yin-Huang-2011}.

\begin{figure}[h]
\centering {\includegraphics[clip,width=10cm]{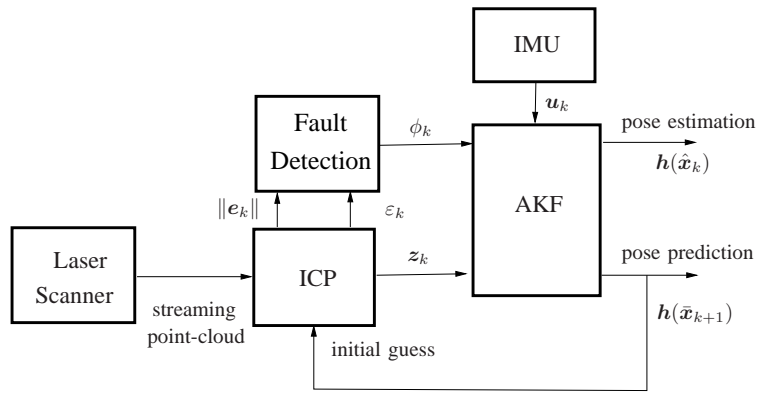}}
\caption{Architecture of the robust relative navigation.}
\label{fig:ekf_lcs}
\end{figure}

It is known that the registration algorithms are not guaranteed to
converge to a global minimum unless the initial alignment is
sufficiently close to the actual one. More specifically,
convergence of ICP iteration and the accuracy of the
fine-alignment  process depend  on quality of the 3-D vision data
that can be adversely affected by many factors such as sensor
noise, disturbance, outliers, symmetric view of the target, or
incomplete scan data. Taking advantage of the simple dynamics of a
free-floating object, which is not acted upon by any external
force or moment, researchers have employed different observers to
track and predict the motion of free-floating space objects
~\cite{Hillenbrand-Lampariello-2005,Aghili-Parsa-2008b,Aghili-Kuryllo-Okouneva-English-2010a}.
However, relative thrust acceleration was not accounted for in
these methods and therefore they are not applicable for relative
navigation. In this work, a relative navigation is developed by
incorporating the dynamics model of the relative pose in the ICP
algorithm to improve the accuracy, performance, and robustness of
the overall pose estimation process. Measurement from onboard IMU
accounts for relative thrust acceleration making the pose
estimation suitable for relative navigation applications. As
depicted in Fig.~\ref{fig:ekf_lcs}, robustness in pose tracking is
achieved by integration of ICP and an adaptive Kalman filter using
a fault-detection logic. Here, the state estimation is  updated
with the value of the fine-alignment as soon as the ICP succeeds
to match the points, while the predicted pose obtained at the
propagation estimate step of the filter gives the coarse alignment
required to find the corresponding points.  A recursive algorithm
updates the covariance estimate associated with the fine-alignment
error at every time step. This information along with detection of
ICP convergence by the fault-detection logic are used to
adaptively tune the Kalman filter for reliable estimation of the
motion states together with a set of relevant parameters, e.g.,
IMU bias and spacecraft CoM. It is proven experimentally that this
closed-loop ICP-AKF architecture, in which the ICP initial guess
is continually provided by the dynamic predictor, establishes
robust pose tracking and allows automatic fault recovery.
Moreover, the computation time can be improved because of the better
initial guess and hence less ICP iterations would be usually
required. The robustness and accuracy performance of the relative
navigation are tested in a robotic experimental setup. In it, the
mock-up of a spacecraft attached to a robotic arm, which is driven
by a simulator to produce the relative motion dynamics between two
objects in orbit. The spacecraft mockup is then scanned by a laser
range sensor while the mockup is in motion and subsequently 3-D
scan data is fed to our relative navigation system in real-time.

\section{Dynamics Model}
Consider a target spacecraft,  namely
the Space Station, and a docking spacecraft and as rigid bodies
moving in nearby orbits. We assume that a laser scanner sensor and
an IMU are installed on the docking vehicle so that they provide
3-D range data of the preceding spacecraft as well as the
acceleration thrust during orbital maneuvers performed in a space
rendezvous and docking. Coordinate frames $\{{\cal A}\}$ and
$\{{\cal B}\}$ are attached to bodies of the docking and target
spacecraft, respectively. The origins of frames $\{{\cal A}\}$ and
$\{{\cal B}\}$ are located at CoMs of the corresponding
spacecraft. Coordinate frame $\{{\cal B}\}$ is orientated so that
its $y$-axis is parallel to a line connecting the Earth's center
to the spacecraft CoM and pointing outward, and its $x$-axis lies
on the orbital plane.  Without loss of generality, we assume that
the coordinate frame, $\{ {\cal A} \}$, coincides with the
coordinate frame of the on-board IMU.

The relative attitude of the docking spacecraft with respect  to
the target spacecraft, which is called the Station hereafter, is
presented by quaternion $\bm q$, but the attitude-control system
of the Station makes it rotate with the angular velocity of the
reference orbit. The rotation matrix $\bm A$ corresponding to the
quaternion $\bm q$ is given by the following expression
\begin{equation} \label{eq:R}
\bm A(\bm q) = (2q_o^2-1) \bm I + 2q_o [\bm q_v \times] + 2 \bm
q_v \bm q_v^T,
\end{equation}
where $\bm q_v$ and $q_o$ are the vector and scaler parts of the
quaternion, i.e., $\bm q=[\bm q_v^T \; q_o]^T$, $[\cdot \times]$ denotes the matrix form of the cross-product,
and $\bm I$ denotes  identity matrix with adequate dimension. The
quaternion product $\otimes$ is defined as
\begin{equation*}
\bm q\otimes= q_o \bm I + \bm\Omega(\bm q_v) \quad \mbox{where}
\quad \bm\Omega(\bm q_v)=\begin{bmatrix} -[\bm q_v \times] & \bm q_v
\\ -\bm q_v^T & 0 \end{bmatrix}
\end{equation*}
so that  $\bm q_1 \otimes \bm q_2$ corresponds to rotation matrix
$\bm A(\bm q_2) \bm A(\bm q_1)$.

Assume that displacement vectors $\bm\varrho_1$ and
$\bm\varrho_2$,  respectively, represent location of the laser
camera in the docking spacecraft w.r.t. its CoM and location of
the point of interest on the Station w.r.t. its own CoM and that
the displacement vectors are expressed in their corresponding
body-attached frames. Also, let us define position vectors
$\bm\rho$ and $\bm r$ as the range of the Station w.r.t. the
docking spacecraft expressed in $\{ {\cal A} \}$ and the relative
distance between the CoMs of two spacecraft expressed in  $\{
{\cal B} \}$, respectively. Then, the following kinematics relation
is in order
\begin{equation} \label{eq:rho}
\bm\rho = \bm A^T (\bm r - \bm\varrho_2) + \bm\varrho_1,
\end{equation}
where $\bm A$ is the rotation matrix from $\{ \mathcal{A} \}$ to
$\{ \mathcal{B} \}$. Considering the visiting spacecraft in the
vicinity of the Station, one can describe the evolution of the
relative position of the two spacecraft by  {\em orbital
mechanics} \cite{Clohessy-Wiltshire-1960}. The thrust-acceleration
measurements of the docking spacecraft $\bm a$ are provided by
IMU. That is
\begin{equation}
\bm a = \bm u_a + \bm b_a + \bm\epsilon_a
\end{equation}
where $\bm u_a$ is acceleration measurement output, $\bm b_a$ is
the bias of the accelerometer, and $\bm\epsilon_a$ is the
accelerometer noise, which is assumed to be random walk noise with
covariance $E[\bm\epsilon_a \bm\epsilon_a^T]=\sigma_a^2 \bm I$. As
shown in the figure, $\bm r_e$ is the vector position connecting
the Earth center to the Station's CoM that is expressed in the
frame  $\{ {\cal B}\}$. Suppose vector $\bm n=n \bm k$, where $\bm
k=[0 \; 0 \; 1]^T$, denote the angular rate of the Station orbit
and hence the angular rate of the rotating frame $\{ {\cal B}\}$.
Then, the relative translational motion of the CoMs  expressed in
frame $\{ {\cal B}\}$ can be described by
\begin{align} \label{eq:ddot_r} 
\ddot {\bm r} =& -2 \bm n \times \dot{\bm r} - \bm n \times(\bm n
\times \bm r) - \dot{\bm n} \times \bm r \\ \notag & + \mu
\Big(\frac{\bm r_e}{\| \bm r_e \|^3}-  \frac{\bm r_e + \bm r}{\|
\bm r_e + \bm r \|^3} \Big) + \bm{A}(\bm
q)(\bm u_a + \bm b_a + \bm\epsilon_a),
\end{align}
where $\mu=3.98\times 10^{14}$~m$^3$/s$^{2}$ is the gravitational parameter of the Earth.

Suppose underlined form of vector
$\bm\alpha\in{\mathbb R}^3$ denotes the representation of that vector in
${\mathbb R}^4$, e.g., $\underline {\bm\alpha}^T \triangleq [\bm\alpha^T \;
0]$. By virtue of this notation, the time-derivative of the quaternion can be expressed by
\begin{equation} \label{eq:dotq}
\dot {\bm q} = \frac{1}{2} \underline{\bm\omega}_{\rm rel} \otimes
\bm q,
\end{equation}
where ${\bm\omega}_{\rm rel}=\bm\omega -\bm\omega_o$ is the
relative angular  velocity,  $\bm\omega_o$ is the angular velocity
of the station expressed in frame $\{{\cal B}\}$, and  $\bm\omega$
is the actual angular rate of the docking spacecraft. The
expression of the latter angular velocity is given by
\begin{equation}
\bm\omega = \bm u_g + \bm b_g + \bm\epsilon_g
\end{equation}
where $\bm u_g$ is the output of gyro,  vector $\bm b_g$ is  the
gyro bias, and $\bm\epsilon_g$ is the gyro noise  with covariance
$E[\bm\epsilon_g \bm\epsilon_g^T]=\sigma_g^2 \bm I$. The
time-derivative of gyro bias is traditionally modeled with random
walk model \cite{Lefferts-Markley-Shuster-1982,Pittelkau-2001}
according to
\begin{equation}\label{eq:dotb_g}
\dot {\bm b}_g = \bm\epsilon_{b},
\end{equation}
where $\bm\epsilon_{b}$ is the rate random walk noise with
covariance $E[\bm\epsilon_{b} \bm\epsilon_b^T]=\sigma_b^2 \bm
I_3$. Furthermore, one  can relate $\bm n$ and $\bm\omega_o$ using
the quaternion transformation \cite{Wilcox-1967} as follow
\begin{equation} \label{eq:omeg_rel}
\underline{\bm\omega}_o = \bm q \otimes \underline {\bm n} \otimes
\bm q^*.
\end{equation}
where $\bm q^*$ is the conjugate quaternion of $\bm q$ such that
$\bm q^* \otimes \bm q = \bm q \otimes \bm q^* = [0 \; 0 \; 0 \; 1
]^T$. Then substituting \eqref{eq:omeg_rel} into \eqref{eq:dotq}
and using the properties of the quaternion products, we arrive at
\begin{align} \notag
\dot {\bm q} &= \frac{1}{2} \underline {\bm\omega} \otimes \bm q -
\frac{1}{2} (\bm q \otimes \underline {\bm n} ) \otimes (\bm q^*
\otimes \bm q)\\ \notag & = \frac{1}{2}  \underline{\bm\omega}
\otimes   \bm q - \frac{1}{2} \bm q \otimes \underline{\bm n} \\
\label{eq:dot_q_nonlin} & = \frac{1}{2}  \bm\Omega(\bm u_g + \bm
b_g + \bm\epsilon_g) - \frac{1}{2} \bm q \otimes \underline{\bm n}
\end{align}

Now, let us suppose that the parameters remain constant during the estimation process, i.e.,
\begin{equation} \label{eq:dot_param}
\dot{\bm b}_a = \dot{\bm\varrho}_1 = \dot{\bm\varrho}_2 = \bm 0
\end{equation}
Then, equations \eqref{eq:ddot_r}, \eqref{eq:dot_param}, and
\eqref{eq:dot_q_nonlin}  can be combined in the following standard
form
\begin{equation} \label{eq:dotxf}
\dot{\bm x} = \bm f(\bm x, \bm u, \bm\epsilon)
\end{equation}
in which vector $\bm u =[\bm u_a^T\; \bm u_g^T]^T$ contains the
IMU outputs, vectors $\bm\epsilon=[ \bm\epsilon_a^T\;
\bm\epsilon_g^T \; \bm\epsilon_b^T]^T$  contains the entire
process noise, and
\begin{equation} \notag
\bm f (\bm x, \bm u, \bm\epsilon) = \begin{bmatrix} \dot{\bm r} \\ -2 \bm n \times \dot{\bm r}  +
\bm\psi(\bm r) + \bm{A}(\bm q)(\bm u_a  + \bm b_a + \bm\epsilon_a) \\
\frac{1}{2} \bm\Omega({\bm u}_g + \bm b_g +  \bm\epsilon_g )  \bm q - \frac{1}{2} \bm q \otimes \underline{\bm n} \\
\bm\epsilon_b\\
\bm 0_{9 \times 1} \end{bmatrix}
\end{equation}
Here, the nonlinear function $\bm\psi(\bm r)$ arises from relative
acceleration due to effects of orbital mechanics and rotating
frame
\begin{equation} \notag
\bm\psi(\bm r) = \mu \Big( \frac{\bm r_e}{\| \bm r_e \|^3}  - \frac{\bm r + \bm r_e }{\|\bm r + \bm r_e \|^3} \Big) - \bm n \times (\bm n \times \bm r)- \dot{\bm n} \times \bm r
\end{equation}
Note that the gyro and accelerometer measurement noises render
the entire process noise and therefore $E[\bm\epsilon
\bm\epsilon^T]=\bm\Sigma_{\rm IMU}= \mbox{diag}( \sigma_{a}^2 \bm
I, \sigma_g^2 \bm I, \sigma_b^2 \bm I)$.

\subsection{Linearization}

Linearized dynamic model is predominantly used in a variety of
navigation systems  across both scientific and engineering realms,
especially in aerospace. Relative navigation estimators are
typically built  upon linearized models, e.g., the extended Kalman
filter  and, to a much lesser extent, the unscented Kalman filter
\cite{Zanetti-2012}. This is because, in practice, it is
sufficient to capture the effects of the relatively small
linearization error as a process noise in the state estimation
process \cite{Valappil-Georgakis-2000}.

Assuming $\| \bm r \| \ll \| \bm r_e \|$, one can obtain
approximating equation of  the orbital mechanics that is {\em
linear} in $\bm r$, i.e.,
\begin{equation}
\bm\psi(\bm r) \approx \bm\Gamma \; \bm r
\end{equation}
where
\begin{align} \notag
\bm\Gamma & = \frac{\partial \bm\psi}{\partial \bm r} \Big|_{\bm r
=0} = -  \frac{\mu}{\| \bm r_e \|^3} \Big(\bm I - 3 \frac{ \bm
r_e \bm r_e^T }{\| \bm r_e \|^2}  \Big) - [\bm n \times]^2 -
[\dot{\bm n} \times] \\ \label{eq:Gamma} & = \frac{\mu}{\| \bm r_e
\|^3} \big( 2 \bm I + 3[\bm j \times]^2 \big)  - n^2[\bm k
\times]^2 - \dot n [\bm k \times]
\end{align}
and $\bm j=\bm r_e/ \| \bm r_e \|$ is the direction cosine. For a
{\em circular}  orbit we have $\dot{\bm n}=0$ and the pertinent
equation becomes
\begin{equation} \notag
n^2 r_e^3 = \mu
\end{equation}
Furthermore, if  we choose coordinate frame $\{ {\cal B} \}$  such
that $\bm j=[0 \; 1 \; 0]^T$, then expression of the jacobian
matrix in \eqref{eq:Gamma} is greatly simplified to
\begin{equation} \notag
\bm\Gamma = \mbox{diag}\big( 0, \; 3n^2, \; -n^2 \big)
\end{equation}

Although the states can be propagated by solving the nonlinear
dynamics equations \eqref{eq:dotxf}, the state transition matrix
of the linearized dynamics equations will be also needed to be
used for covariance propagation of the KF. Since additive error
quaternion does not lead to  a unit vector, a quaternion variation
$\tilde{\bm q}$ using a multiplication error quaternion involving
{\em a posterior} quaternion $\bar{\bm q}$ is considered for state
estimation so that $\tilde{\bm q} \otimes \bar{\bm q} = \bm q$, or
equivalently
\begin{equation} \label{eq:delta_q}
\tilde{\bm q}= \bm q \otimes \bar{\bm q}^*.
\end{equation}
Assuming small $\bm b_g$ and $\bm\epsilon_g$, one can adopt a linearization technique
similar to~\cite{Lefferts-Markley-Shuster-1982,Pittelkau-2001} to linearize \eqref{eq:dotq} about the  quaternion
$\bar{\bm q }$ and angular velocity $\hat{\bm\omega}= \hat{\bm u}_g + \hat{\bm b}_g$
\begin{equation} \label{eq:diff_delqv}
\frac{d}{dt} \tilde{\bm q}_v = - \hat{\bm\omega} \times \tilde{\bm q}_v +
\frac{1}{2} \big( \bm b_g  - \hat{\bm b}_g)  + \frac{1}{2} \bm\epsilon_{g}
\end{equation}
Since $\tilde q_o$ is not an independent variable and it has
variations of only the second order, its time-derivative can be
ignored, as suggested in~\cite{Lefferts-Markley-Shuster-1982}.
Moreover, it can be seen from \eqref{eq:R} that for small rotation
$\tilde{\bm q}$,  where $\| \tilde{\bm q}_v \| \ll 1 $, we have
$\bm A(\tilde{\bm q} ) = \bm I + 2[\tilde{\bm q}_v \times]$ and
thus
\begin{equation} \label{eq:A(deltaq)}
\bm A(\bm q)  \approx \bar{\bm A} \big( \bm I + 2[\tilde{\bm q}_v \times]  \big)
\end{equation}
where $\bar{\bm A} = \bm A(\bar{\bm q})$. Now the equation of
translational motion \eqref{eq:ddot_r} can be linearized about the
acceleration estimation $\hat{\bm u}_a + \hat{\bm b}_a$ and  {\em
a posterior} quaternion $\bar{\bm q}$  using the first-order
approximation  \eqref{eq:A(deltaq)} as:
\begin{align} \notag
\ddot{\bm r} &  \approx -2 \bm n \times \dot{\bm r} + \bm\Gamma
\bm r -2 \bar{\bm A}  \big((\hat{\bm u}_a + \hat{\bm b}_a) \times
\tilde{\bm q}_v \big)  \\ \label{eq:diff_delr} &+ \bar{\bm A}
\big( \bm b_a - \hat{\bm b}_a)  + \bar{\bm A} \bm\epsilon_a .
\end{align}
In order to avoid introducing new variable, let us redefine
\begin{equation}
\bm x=[\bm r^T \; \dot{\bm
r}^T \; \tilde{\bm q}_v^T \; \bm b_g^T \; \bm b_a^T \; \bm\varrho_1^T \; \bm\varrho_2^T]^T,
\end{equation}
to be the state vector of the KF estimator. Then, setting
\eqref{eq:dotb_g}, \eqref{eq:diff_delqv}, and \eqref{eq:diff_delr}
in the state-space form, the linearized model of the continuous
system can be derived as
\begin{subequations} \label{eq:linmodel}
\begin{equation}
\delta \dot{\bm x} = \bm F \delta \bm x + \bm G \bm\epsilon
\end{equation}
where
\begin{align}\label{eq:F}
\bm F & =
\begin{bmatrix}
\bm 0 & \bm I & \bm 0 & \bm 0 & \bm 0 & \bm 0\\
\bm\Gamma &  -2n[\bm k \times]  & -2 \bar{\bm A}[(\hat{\bm u}_{a} + \hat{\bm b}_{a}) \times] &  \bar{\bm A} & \bm 0 & \bm 0\\
\bm 0 & \bm 0 & -[(\hat{\bm u}_{g} + \hat{\bm b}_{g}) \times ] & \bm 0 & \frac{1}{2} \bm I & \bm 0\\
\bm 0 & \bm 0 & \bm 0 & \bm 0 & \bm 0 & \bm 0
\end{bmatrix}\\ \label{eq:G}
\bm G &  =\begin{bmatrix}
\bm 0 & \bm 0 & \bm 0\\
\bar{\bm A} & \bm 0 & \bm 0 \\
\bm 0 & \frac{1}{2} \bm I & \bm 0\\
\bm 0 & \bm 0 & \bm I\\
\bm 0 & \bm 0 & \bm 0
\end{bmatrix}
\end{align}
\end{subequations}
The equivalent discrete-time system of the linearized system can
be written as
\begin{equation} \label{eq:discere_sys}
\delta \bm x_{k+1} = \bm\Phi_k  \delta \bm x_k + \bm w_k,
\end{equation}
where $\bm w_k$ is discrete-time process noise, and the state
transition matrix $\bm\Phi_k = \bm\Phi(t_k, t_{\Delta})$ at time
$t_k$ over time interval  $t_{\Delta} = t_{k+1}-t_k$ is given by
\begin{equation} \label{eq:Phi}
\bm\Phi(t_k, t_{\Delta}) = e^{\bm F(t_k) t_{\Delta}} \approx \bm I +  t_{\Delta} \bm F(t_k).
\end{equation}
The covariance of the discrete-time process noise is related to the
continuous process noise covariance by
\begin{equation} \label{eq:Q_integral}
\bm Q_k=E[\bm w_k \bm w_k^T] = \int_{t_k}^{t_k+t_{\Delta}}
\bm\Phi(t_k,\tau) \bm G\bm\Sigma_{\rm IMU} \bm G^T \bm\Phi^T(t_k,\tau) {\rm d}
\tau,
\end{equation}
Subsequently using the first-order approximation of the
exponential matrix from \eqref{eq:Phi}  together with the
expression of $\bm F$ and $\bm G$ from \eqref{eq:F} and
\eqref{eq:G} into \eqref{eq:Q_integral}  and then carrying out
tedious calculation of the integral, we arrive at the expression
of the covariance matrix in the following form
\begin{subequations} \label{eq:Q_expression}
\begin{equation}
\bm Q_{k} = \begin{bmatrix} \bm Q_{k_{11}} & \bm Q_{k_{12}} & \bm 0 & \bm 0 & \bm 0 \\
\times & \bm Q_{k_{22}} & \bm Q_{k_{23}} & \bm Q_{k_{24}} & \bm 0  \\
\times & \times & \bm Q_{k_{33}} & \bm 0 & \bm 0\\
\times & \times & \times & \bm Q_{k_{44}} & \bm 0 \\
\times & \times & \times & \times & \bm 0
\end{bmatrix},
\end{equation}
where the sub-matrices are
\begin{align}
\bm Q_{k_{11}} & = \frac{1}{3} \sigma_a^2 t_{\Delta}^3 \bm I\\
\bm Q_{k_{12}} & = \sigma_a^2 \big( \frac{1}{2} t_{\Delta}^2 \bm I + \frac{2}{3} t_{\Delta}^3 n [\bm k \times] \big)\\
\bm Q_{k_{22}} & = \sigma_a^2 \big( t_{\Delta} \bm I - \frac{4}{3} t_{\Delta}^3 n^2 [\bm k \times]^2 \big) + \frac{1}{3} \sigma_b^2 t_{\Delta}^3 \bm I\\ \notag
& - \frac{1}{3} \sigma_g^2 t_{\Delta}^3 \bar{\bm A}_k[(\hat{\bm u}_{a_k} + \hat{\bm b}_{a_k}) \times]^2 \bar{\bm A}_k^T \\
\bm Q_{k_{23}} & = \sigma_g^2 \bar{\bm A}_k \Big( \frac{1}{6} t_{\Delta}^3 [(\hat{\bm u}_{a_k} + \hat{\bm b}_{a_k}) \times][(\hat{\bm u}_{g_k} + \hat{\bm b}_{g_k}) \times]  \\ \notag
& + \frac{1}{4} t_{\Delta}  [(\hat{\bm u}_{a_k} + \hat{\bm b}_{a_k}) \times] \Big)\\
\bm Q_{k_{24}} & = \frac{1}{4} \sigma_b^2 t_{\Delta}^2  \bar{\bm A}_k\\
\bm Q_{k_{33}} & = \sigma_g^2 \big( \frac{1}{4} t_{\Delta} \bm I - \frac{1}{12} t_{\Delta}^3 [(\hat{\bm u}_{g_k} + \hat{\bm b}_{g_k}) \times]^2  \big)\\
\bm Q_{k_{44}} & = \sigma_b^2 t_{\Delta} \bm I.
\end{align}
\end{subequations}
Notice that since the covariance matrix is symmetric, the
sub-matrices in lower triangular part  of $\bm Q_k$, which are
denoted by the cross signs,  are transposes of the corresponding
sub-matrices in the upper triangular part. The process noise
covariance matrix can be systematically calculated in
\eqref{eq:Q_expression} from where  a priori knowledge of the IMU
noise parameters   $\{\sigma_a, \; \sigma_g, \; \sigma_b\}$, which
are usually specified by the manufacturer. However, we treat them
as the filter's "tuning parameters" to capture the linearization
errors so that confidence in the linearized model can be properly
specified \cite{Valappil-Georgakis-2000}.

Alternatively, the exact solution to the state transition matrix
and discrete-time process noise  can be obtained from the van Loan
method \cite{vanLoan-1978}  through the following steps: i)
Construct  upper triangular matrix $\bm\Lambda$ from  \eqref{eq:F}
and \eqref{eq:G} and then compute the exponential matrix $\bm\Psi$
\begin{equation} \notag
\bm\Lambda \triangleq \begin{bmatrix} -\bm F &
\bm G \bm\Sigma_{\rm IMU} \bm G^T \\ \bm 0 & \bm F^T
\end{bmatrix} \quad,   \qquad \bm\Psi=e^{\bm\Lambda t_{\Delta}}
\end{equation}
with $t_{\Delta}$ being the sampling time; ii) partition the
resultant upper-triangular matrix  $\bm\Psi$ into sub-matrices
$\bm\Psi_{11}$, $\bm\Psi_{12}$, and $\bm\Psi_{22}$ for
construction of $\bm\Phi_k$  and $\bm Q_k$
\begin{equation} \label{eq:vanLoan}
\bm\Psi = \begin{bmatrix} \bm\Psi_{11} & \bm\Psi_{12} \\ \bm 0 & \bm\Psi_{22} \end{bmatrix} \quad \Longrightarrow \quad
\bm\Phi_k=\bm\Psi_{22}^T, \quad \bm Q_k= \bm\Psi_{22}^T\bm\Psi_{12}
\end{equation}
Note that although \eqref{eq:vanLoan} yields more accurate results
than  formulation  \eqref{eq:Phi}-\eqref{eq:Q_expression}, the
latter does not require computation of exponential matrix and
hence it is superior numerically to the former.

\subsection{Observation Equations}

As will be discussed later in Section~\ref{sec:ICP}, a noisy
measurement of  the relative pose can be obtained from
registration of two 3-D point sets in the ICP algorithm. Assume
that ${\bm\rho}'$ and ${\bm q}'$, respectively, represent the
position and orientation parts of the rigid-body transformation at
the end of the fine-alignment step of the data registration. Then
similar to \eqref{eq:delta_q}, one can  transform the measured
quaternion into small quaternion variation $\tilde{\bm q}'={\bm
q}' \otimes \bar{\bm q}^{*}$, and
\[\tilde{\bm q}'_v = \bm\Lambda(\bar{\bm q}) {\bm q}', \quad \mbox{where} \quad \bm\Lambda(\bar{\bm q}) = \begin{bmatrix} {\bar q}_o \bm I -[\bm \bar{\bm q}_v \times] & - \bar{\bm q}_v \end{bmatrix} \]
Therefore, the observation vector is given by
\begin{equation}
\bm z= \begin{bmatrix} {\bm\rho}' \\ \bm\Lambda(\bar{\bm q}) {\bm
q}' \end{bmatrix}.
\end{equation}
On the other hand, the observation equation can be written as a function of the state vector
\begin{equation}
\bm z  =  \bm h(\bm x) + \bm v,
\end{equation}
where $\bm v$ is the observation noise and the corresponding
covariance $E[\bm v \bm v^T]=\bm R$ is  allowed  to be
time-varying. The observation variables can be expressed in terms
of the states variables as follows
\begin{equation} \label{eq:h_nonlin}
\bm h(\bm x) = \begin{bmatrix}  \bm A^T( \tilde{\bm q} \otimes \bar{\bm q}) (\bm r -\bm\varrho_2) + \bm\varrho_1   \\ \tilde{\bm q}_v
\end{bmatrix}.
\end{equation}
Note that the observation vector \eqref{eq:h_nonlin} is a
nonlinear function of the states. In view of \eqref{eq:A(deltaq)},
we can linearize the range measurement  equation \eqref{eq:rho} in
terms of $\tilde{\bm q}$ as follow
\begin{equation} \label{eq:r}
\bm\rho \approx  \big( \bm I - 2 [\Tilde{\bm q}_v \times]
\big)\bar{\bm A}^T (\bm r - \bm\varrho_2) + \bm\varrho_1.
\end{equation}
Therefore, we can write the sensitivity matrix as
\begin{align} \label{eq:H}
\bm H_k &= \left. \frac{\partial \bm h}{\partial \bm x} \right|_{\bar{\bm x}_k} \\ \notag
&=\begin{bmatrix} \bar{\bm A}_k^T & \bm 0 & 2[(\bar{\bm A}_k^T (\bar{\bm r}_k - \bar{\bm\varrho}_{2_k})) \times] & \bm 0 & \bm 0 & \bm I & - \bar{\bm A}_k^T\\
\bm 0 & \bm 0 & \bm I & \bm 0 & \bm 0 & \bm 0 & \bm 0
\end{bmatrix}
\end{align}
Now, with the state transition matrix and the observation
sensitivity matrix in hand, one can design the estimator as
elaborated in Appendix~A.

\subsection{Estimation of the Observation Covariance Matrix} \label{sec:noise-adaptive}

Efficient implementation of the KF requires the statistical
characteristics of observation error in the innovation sequence of
the filter \cite{Aghili-Parsa-2008b}. The observation  error is related to the pose
refinement step of ICP that depends on not only the quality of the
3D data acquired by the laser scanner, e.g., noise and outliers,
but also the performance of ICP to converge to global minima.
Unexpected noise and disturbance in the laser range sensor is
largely responsible for the ICP error and thus a priori knowledge
of the corresponding covariance is not usually available.
Therefore, in order to improve the quality of the pose estimate
requires weighting the ICP output with the proper data more
heavily than the one with ``poor'' data in the estimator; rather
than giving all ICP outputs equal weights. This goal can be
achieved by readjusting the observation covariance matrix  in the
filter's internal model, so that the filter is tuned as much
as possible. In a noise-adaptive Kalman filter, the issue is that,
in addition to the states, the covariance of the observation noise
has to be estimated~\cite{Chui-Chen-1998-p113}. Adaptive Kalman
filter based on  {\em maximum likelihood estimation} was
originally proposed by Mehra \cite{Mehra-1970} followed by
derivation of several variations and investigation of stability
analysis
\cite{Wang-2000,Kim-Jee-Park-Lee-2009,Gao-Wei-Zhong-Subic-2015}.
The covariance matrix of the observation can be obtained from
averaging the sequence of either the innovation matrix or the
residual matrix  inside a moving window of size $w$. Let us define
the {\em innovation sequence} ${\bm e}_k$ as the difference
between incoming pose update from ICP loop $\bm z_k$ and the
predicted pose obtained from the {\em a priori} state estimate
$\bar{\bm x}_{k}$, i.e.,
\begin{subequations}
\begin{align} \label{eq:ek}
{\bm e}_k &= \bm z_k - \bm H_k \bar{\bm x}_k \\\label{eq:varrho}
&= \bm H_k({\bm x}_k - \bar{\bm x}_k) + \bm v_k.
\end{align}
\end{subequations}
If we assume the  estimation process remains constant over the
most recent $w$ steps, we can  arrive estimate at estimation of
the observation covariance  by  taking variance of both sides of
\eqref{eq:varrho}
\begin{equation} \notag
\frac{1}{w}\sum_{i=1}^{w} {\bm e}_{k-i}
{\bm e}_{k-i}^T =  \bm H_{k} \bar{\bm P}_k \bm H_{k}^T   +  \hat{\bm R_k}
\end{equation}
or
\begin{subequations}  \label{eq:R_estimate1}
\begin{equation} \label{eq:R_innovation}
\hat{\bm R_k} \approx \hat{\bm C}_k - \bm H_k \bar{\bm P}_k \bm H_k^T
\end{equation}
where
\begin{equation} \label{eq:S_batch}
\hat{\bm C}_k  = \frac{1}{w}\sum_{i=k-w}^{k-1} {\bm e}_i
{\bm e}_i^T
\end{equation}
\end{subequations}
is the {\em innovation covariance matrix} averaged inside the
sliding estimation window, and  covariance of the predicted state
vector $\bar{\bm P}_k=E[({\bm x}_k - \bar{\bm x}_k)({\bm x}_k -
\bar{\bm x}_k)^T]$ is assumed constant within the window.

An alternative estimate of the observation covariance can be
derived from the residual sequence  instead of the innovation
sequence. Let us define the {\em residual sequence}
\begin{equation}
{\bm e}_k^* = \bm z_k - \bm H_k \hat{\bm x}_k
\end{equation}
as the difference between the observed pose and the estimated one
obtained from {\em a priori}  state estimate $\hat{\bm x}_k$.
Then, it can be shown  that expression of the estimated
observation covariance takes the following form \cite{Wang-2000}
similar to \eqref{eq:R_innovation}
\begin{subequations}  \label{eq:R_estimate2}
\begin{equation} \label{eq:R_resudua2}
\hat{\bm R_k} \approx \hat{\bm C}^*_k + \bm H_k \hat{\bm P}_k \bm
H_k^T
\end{equation}
where the  {\em innovation covariance matrix} is:
\begin{equation} \label{eq:S_batch2}
\hat{\bm C}^*_k  = \frac{1}{w}\sum_{i=1}^{w} {\bm e}^*_{k-i}
{\bm e}_{k-i}^{*T}
\end{equation}
\end{subequations}

Either formulation \eqref{eq:R_estimate1} or
\eqref{eq:R_estimate2} can be used for estimation of the
observation covariance matrix $\hat{\bm R}_k$ from an {\em
ergodic} approximation of the covariance of the corresponding
error in the sliding sampling window with finite length $w$.
Nevertheless, only expression  \eqref{eq:R_estimate2} guarantees
that the outcome observation covariance matrix remains always
positive-definite \cite{Aghili-Parsa-2009}. It should be pointed out that  $w$ has to be
chosen empirically to give some statistical smoothing. The
intuitive motion in choosing a finite window in the estimation of
the innovation covariance matrix is that very past error data has
to be discounted when being used for estimation of the current
covariance. If the innovation covariance is not expected to change
significantly over time, then a large sample size can be chosen
for accurate estimation of the covariance matrix, otherwise the
window should be selected as small as possible.

The recursive version of the innovation-based or the
residual-based covariance  estimation can be derived as following
\begin{equation} \label{eq:S_recursive}
\hat{\bm C}_{k+1} = \left\{ \begin{array}{ll} \frac{k-1}{k}\hat{\bm
C}_{k} + \frac{1}{k} \bm e_{k}  \bm e_{k}^T & \quad \text{if} \quad k<w\\
\hat{\bm C}_{k} + \frac{1}{w} \Big( \bm e_{k} \bm e_{k}^T -
\bm e_{k-w}\bm e_{k-w}^T \Big) & \quad \text{otherwise}
\end{array} \right.
\end{equation}
Note that  $\hat{\bm C}$ and $\bm e$ can be replaced by $\hat{\bm
C}^*$ and $\bm e^*$ in \eqref{eq:S_recursive}  for recursive
representation of the residual-based method.  The following block
diagram illustrates the realization of the above recursive
estimator for on-line estimation of the innovation covariance
matrix.
\begin{figure}[h]
\centering {\includegraphics[clip,width=10cm]{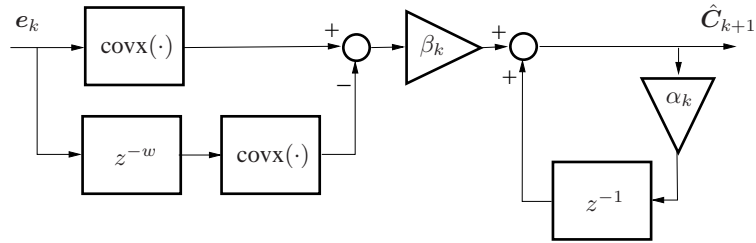}}
\caption{Online estimation of the  innovation covariance.}
\label{fig:cov_estim}
\end{figure}
Here, $\mbox{covx}(\bm a)=\bm a \bm a^T$, $z^{-1}$ block
represents {\em unit delay} and hence  $z^{-w}$ block represents
$w$ cascaded unit delay, and time-varying scaler variables
$\alpha_k$ and $\beta_k$ are defined as
\begin{equation} \notag
\alpha_k = \left\{ \begin{array}{ll} \frac{k-1}{k} & \quad \text{if} \quad k<w\\
1 & \quad \text{otherwise}
\end{array} \right. , \quad  \beta_k = \left\{ \begin{array}{ll} \frac{1}{k} & \quad \text{if} \quad k<w\\
\frac{1}{w} & \quad \text{otherwise}
\end{array} \right.\end{equation}
We assume that the initial states of the memory block $z^{-w}$ are
zero while the output of  memory block $z^{-1}$ is specified as
initial guess $\bm C_0$ for the first time.

\section{Registration of two 3-D data points} \label{sec:ICP}

Although the registration process of 3D CAD model and point cloud
has been used by 3D vision algorithms for more than two decades,
robust registration of  two 3D point sets is still a common
problem in computer vision \cite{Shang-Jasiobedzki-2005}.  The
iterative closest point (ICP) algorithm is  the most popular
algorithm for solving this kind of problem. Although there are
many variants of the ICP algorithm, the basic ICP estimates the
pose of an object in two steps as follows:

{\em Step I: Find the corresponding points on the model set
assuming a coarse alignment is given.} Suppose that we are given
with a set of 3-D points data ${\cal C}$ that corresponds to a
single shape represented by {\em model set} ${\cal M}$, which is
usually a CAD model. It is known that for each point $\bm
c_i\in\mathbb{R}^3$ from the 3-D points data set ${\cal C}=\{\bm
c_1 \cdots \bm c_m \}$, there exists at least one point on the
surface of ${\cal M}$ which is closer to $\bm c_i$ than other
points in ${\cal M}$~\cite{Simon-Herbert-Kanade-1994}. Assume that
the initial rigid  transformation $\{\bm q'(0), \bm\rho'(0) \}$ is
given from prediction of the states at the propagation step of KF
through the followings:
\begin{subequations} \label{eq:initial_pose}
\begin{equation}
\bm q'(0) = \bar{\bm q}_{k}
\end{equation}
\begin{equation} \notag
\bm\rho'(0)= \bm{A}^T(\bar{\bm q}_k)(\bar{\bm r}_k - \bar{\bm\varrho}_{2_k}) + \bar{\bm\varrho}_{k_1}
\end{equation}
\end{subequations}
Then, the problem of finding the correspondence between the two sets can be formally expressed by
\begin{equation} \label{eq:ci}
\bm d_i = \mbox{arg} \min_{\bm d_j \in {\cal M} } \|\bm A(\bm q'(0)) \bm c_i
+ \bm\rho'(0) - \bm d_j \| \qquad \forall i=1,\cdots,m  ,
\end{equation}
and subsequently set ${\cal D} = \{\bm d_1 \cdots \bm d_m \}$ is
formed. Now, we have two independent sets of 3-D points ${\cal C}$
and ${\cal D}$ both of which corresponds to a single shape but
they are related to each other  through a rigid-body
transformation.

{\em Step II: Finding a fine-alignment between two data sets.} The
next problem is to find a fine-alignment $[\bm\rho^T \; \bm
q^T]^T$ which minimizes the distance between these two sets of
points~\cite{Besl-Mckay-1992}. This can be formally formulated as
\begin{align} \label{eq:ICP}
& \varepsilon = \mbox{arg  } \min_{\bm q, \bm\rho} \frac{1}{m} \sum_{i=1}^m \| \bm A(\bm q) \bm c_i +
\bm\rho - \bm d_i \|^2 \qquad \forall \bm d_i \in {\cal D}, \bm c_i
\in {\cal C} \\ \notag & \qquad \text{s.t.} \quad  \| \bm q \| =1
\end{align}
where $\varepsilon$ is called ICP metric fit error. There are
several closed-form  solutions to the above least-squares
minimization problem
\cite{Faugeras-Herbert-1986,Eggert-Lorusso-Fisher-1997}, of which
the quaternion-based algorithm  is the preferable choice for $m>3$
over other methods \cite{Horn-1987,Besl-Mckay-1992}.  Consider the
cross-covariance matrix of the sets ${\cal C}$ and  ${\cal D}$
given by
\begin{equation}
\bm S = \mbox{cov}({\cal C}, {\cal D}) = \frac{1}{m} \sum_{i} \bm
c_i \bm d_i^T -  \bar{\bm c} \bar{\bm d}^T
\end{equation}
where $\bar{\bm c} = \frac{1}{m} \sum_{i} \bm c_i$ and $\bar{\bm d}
= \frac{1}{m} \sum_{i} \bm d_i$ are the corresponding centroid of
the sets. Let us construct the following symmetric weighting
matrix  from cross-covariance matrix as follow
\begin{equation} \notag
\bm W = \begin{bmatrix}
\mbox{tr}(\bm S) & \bm s^T \\
\bm s & \bm S + \bm S^T - \mbox{tr}(\bm S) \bm I
\end{bmatrix},
\end{equation}
where $\bm s=[S_{23} - S_{32}, S_{31} -S_{13}, S_{12}-S_{21}]^T$. Then, it
has shown that minimization problem \eqref{eq:ICP} is tantamount
to find the solution of following quadratic programming
\cite{Horn-1987}
\begin{equation} \label{eq:quadratic}
\max_{\| \bm q \| =1} \bm q^T \bm W \bm q
\end{equation}
As shown in Appendix~B, the solution of the preceding optimization problem is
\begin{equation} \label{eq:q_rime}
\begin{array}{cc}
{\bm q}' &={\mbox{eigenvector}(\bm W)}\\
& \lambda_{\rm max}(\bm W)
\end{array}
\end{equation}
That is  the eigenvector corresponding to the largest eigenvalue
of  matrix $\bm W$. Then, the translation can be obtained from
\begin{equation}  \label{eq:r_rime}
{\bm\rho}' = \bar{\bm d} - \bm A({\bm q}') \bar{\bm c}
\end{equation}
It should be mentioned that there is also a preprocessing step in
the  registration process where the 3D CAD model is converted into
suitable representation for the ICP. The 3D CAD model is often
exported into {\em stereolithography} (STL) format consisting of
normal and triangular facets, which are then used as a biases to
generate a point cloud model with uniform resolution. Generally
speaking, the point cloud obtained from the CAD model should have
comparable resolutions with the one acquired by the laser scanner
in order to maintain accurate registration results
\cite{Li-Yin-Huang-Xiong-2011}. Therefore, re-sampling of model
points may be required as  the resolution  of the scanned point
data set depends on the distance between the scanner sensor and
the object. In our application, the point cloud model is generated
once for use in all ICP cycles for the sake of simplicity of
implementation. Nevertheless, the model points can be subject to a
data re-sampling process using different methods as discussed in
\cite{Kim-Lee-Cho-Kim-2011}.

In the conventional ICP approaches, the incremental transformation
obtained from the fine-alignment step is applied to step I and
then it is iterated until the distance between the two sets of
points in \eqref{eq:ICP} is less than the pre-specified  threshold
$\varepsilon_{\rm th}$. As illustrated  in
Algorithm~\ref{algo:ICp-EKF-alg}, however, the process of our
robust pose estimation algorithm follows these steps: If  the ICP
matching error $\varepsilon$ becomes lower than the threshold
$\varepsilon_{\rm th}$ before the maximum number of iterations
$i_{\rm max}$ is reached, then ICP iteration is considered
convergent. In this case, the rigid-body transformation obtained
from the refinement step, i.e., \eqref{eq:q_rime} and
\eqref{eq:r_rime}, is used not only to correct state estimation in
the KF innovation sequence but to update covariances  of state
estimation and  observation error \eqref{eq:KF_update}.
Subsequently, a posteriori values of the state vector and the
state covariance matrix are computed through the propagation step
\eqref{eq:KF_propagate}. The predicated pose is set as the coarse
alignment to find the corresponding points (step I of ICP) in the
next time step as described in \eqref{eq:initial_pose}. On the
other hand, if $\varepsilon > \varepsilon_{\rm th} $ but the
maximum number of iteration is  not reached, then the  rigid-body
transformation obtained from  \eqref{eq:q_rime} and
\eqref{eq:r_rime} is used to set the initial pose and the
iterations are carried out at the current time step. Finally, if
the maximum number of iteration is reached while the data set
matching error is still not below the  threshold $\varepsilon_{\rm
th}$, then the ICP iteration is considered a failure and
subsequently the {\em a priori} state estimation is used to
compute pose estimate.

Suppose the following flagged variable indicates wether the 3D
registration  process at epoch $k$  is healthy or faulty
\begin{equation}
\phi_k = \left\{ \begin{array}{ll} 0 \quad & \mbox{ICP fault}  \\ 1 & \mbox{Otherwise} \end{array} \right.
\end{equation}
As mentioned earlier, a large ICP metric fit error indicates
non-convergent ICP and  that is at the core of our fault-detection
mechanism. However, even convergent ICP  does not always guarantee
that the pose estimation outcome is valid. This is because even if
the metric error becomes relatively small, it is still possible to
get erroneous results due to multiple effects such as sensor
noise, outliers, or convergence to local minima. The validity of
the ICP pose estimation outcome can be examined by comparing it
with the predicted pose derived from the dynamics model and using
acceleration and rotation data measured by IMU. A significant
mismatch then can be also interpreted as ICP fault. Therefore, the
condition for detecting ICP fault can be expanded as
\begin{equation}
\varepsilon \geq \varepsilon_{\rm th} \quad \mbox{or} \quad  \| \bm e_k \|_W \geq e_{\rm th},
\end{equation}
where $\bm e_k$ is difference between ICP pose estimate and pose
predication  based on  the {\em a priori} state estimate
\eqref{eq:ek}, $e_{\rm th}$  is the pose error threshold, and $\|
\cdot \|_W$ denotes weighted Euclidian norm. Once ICP fault is
detected, the pose-tracking fault recovery   is rather
straightforward. To this end, the Kalman filter gain is set to
\begin{equation} \label{eq:K_phi}
\bm K_k  = \phi_k \bar{\bm P}_k \bm H_k^T \hat{\bm C}_k^{-1}
\end{equation}
Clearly $\bm K_k = \bm 0$ when ICP fault is detected, in which
case the  observation information is not incorporated in the
estimation process  for updating the state and the estimator
covariance, i.e., $\phi_k=0 \; \Longrightarrow \; \hat{\bm x}_k =
\bar{\bm x}_k, \;\; \hat{\bm P}_k = \bar{\bm P}_k$. 
\begin{algorithm}
\DontPrintSemicolon
\KwIn{measurements ${\cal C}_k =\{\bm c_{1_k} ,\cdots, \bm c_{m_k} \}$ , $\bm u_k$}
\KwOut{Pose estimate}
$i \gets 0$\;
\While{$i < i_{\rm max}$}{
     $i \gets i+1$\;
     ICP step I:
     ${\cal D}=\{\bm d_1 , \cdots , \bm d_m \} \gets \bm q'(0), \bm\rho'(0), {\cal C}_k$\;
     ICP step II: $\{ \bm q', \bm\rho', \varepsilon \} \gets {\cal D}, {\cal C}_k$ \;
     \uIf{$\varepsilon \leq \varepsilon_{\rm th} $}
    {Innovation sequence:
     $\{ \hat{\bm q}_k, \hat{\bm\rho}_k, \hat{\bm x}_k \}\gets \{ \bm q', \bm\rho' \}$\;
     Update covariance:  $\hat{\bm C}_k \gets \{ \hat{\bm e}_k \hat{\bm e}^T_k , \hat{\bm C}_{k-1} \}$\;
     Propagation sequence: $\{ \bar{\bm q}_{k+1}, \bar{\bm\rho}_{k+1}, \bar{\bm x}_{k+1} , \bar{\bm P}_{k+1} \}
     \gets \{ \hat{\bm x}_k , \bm u_k, \hat{\bm P}_k \}$ \; $\bm q'(0) \gets \bar{\bm q}_{k+1}
     \quad \wedge \quad \bm\rho'(0) \gets \bar{\bm\rho}_{k+1}$\;
     $\phi_k \gets 1$\;
     \Return{$\hat{\bm q}_k,  \hat{\bm\rho}_k, \phi_k $}
    }
    \Else{
      $\bm q'(0) \gets {\bm q}' \quad \wedge \quad \bm\rho'(0) \gets {\bm\rho}'$ \;
    }
} Update: $\hat{\bm x}_k \gets \bar{\bm x}_k$, $\hat{\bm C}_k
\gets \hat{\bm C}_{k-1}$, $\hat{\bm P}_k \gets \bar{\bm P}_k$\;
Propagation sequence: $\{ \bar{\bm q}_{k+1}, \bar{\bm\rho}_{k+1},
\bar{\bm x}_{k+1}, \bar{\bm P}_{k+1}  \} \gets \{ \hat{\bm x}_k,
\bm u_k , \hat{\bm P}_k \}$\;
     $\bm q'(0) \gets \bar{\bm q}_{k+1} \quad \wedge \quad \bm\rho'(0) \gets \bar{\bm\rho}_{k+1}$\;
$\phi_k \gets 0$\; \Return{$ \hat{\bm q}_k, \hat{\bm\rho}_k,
\phi_k$ }\; \caption{{\sc ICP-AKF Integration}}
\label{algo:ICp-EKF-alg}
\end{algorithm}

\appendix
\subsection*{Appendix A: Noise adaptive Kalman filter} \label{appx:kalman}

The Kalman-based observer for the associated linearized system
\eqref{eq:linmodel}-\eqref{eq:H} is given in the following steps:

\begin{enumerate}
\item  Update:  \\
\begin{subequations} \label{eq:KF_update}
\begin{align} \label{eq:K_est}
\bm K_k & = \phi_k \bar{\bm P}_k \bm H_k^T \hat{\bm C}_k^{-1} \\
\hat{\bm x}_k & = \bar{\bm x}_k + \bm K_k \big(\bm z_k - \bm h(\bar{\bm x}_k) \big) \\ \label{eq:x_est}
\hat{\bm q}_k &= \hat{\tilde{\bm q}}_k \otimes \bar{\bm q}_k\\
\hat{\bm P}_k &= \big( \bm I - \bm K_k \bm H_k  \big) \bar{\bm P}_k
\end{align}
\end{subequations}
\item On-line estimation of the covariance matrix from \eqref{eq:S_recursive}
\item Propagation:
\begin{subequations} \label{eq:KF_propagate}
\begin{align}\label{eq:state-prop}
\bar{\bm x}_{k+1} & = \hat{\bm x}_{k} +
\int_{t_k}^{t_{k}+t_{\Delta}} \bm f(\bm x, \bm u(\tau), \bm
0)\,{\text d} \tau\\ \label{eq:cov-prop} \bar{\bm P}_{k+1}&=
\hat{\bm\Phi}_{k} \bm P_{k} \hat{\bm\Phi}_{k}^T + \bm Q_{k}
\end{align}
\end{subequations}

For non-adaptive KF, the covariance matrix on the innovation can
be constructed  from a given covariance matrix of the vision
system noise by
\begin{equation} \notag
\bm C_k = \bm H_k \bar{\bm P}_k \bm H_k^T + \bm R_k
\end{equation}
\end{enumerate}

\subsection*{Appendix B: Optimal quaternion estimate} \label{appx:opt_quaternion}

Denote the Lagrangian equation of the constrained
quadratic programming \eqref{eq:quadratic} as
\begin{equation} \notag
L = \bm q^T \bm W \bm q - \lambda(\bm q ^T \bm q -1)
\end{equation}
where $\lambda$ is the Lagrangian multiplier. Then, the optimal solution
has to satisfy the stationary condition, which leads to the
following equation:
\begin{equation} \label{eq:lagrangian_derivative}
\frac{\partial L}{\partial \bm q} = 0  \quad \Rightarrow \quad \bm W \bm q - \lambda_{\rm max} \bm q = \bm 0
\end{equation}
Notice that  \eqref{eq:lagrangian_derivative} is indeed equivalent
to the characteristic equation of matrix $\bm W$ for the largest
eigenvalue $\lambda_{\rm max}=\lambda_{\rm max}(\bm W)$.  In other
words, the optimal quaternion is the eigenvector of matrix $\bm W$
corresponding to its largest eigenvalue.

\bibliographystyle{IEEEtran}

\begin{thebibliography}{10}
\bibitem{Aghili-2016c}
F.~Aghili and C.~Y. Su, ``Robust relative navigation by integration of icp and
  adaptive kalman filter using laser scanner and imu,'' \emph{IEEE/ASME
  Transactions on Mechatronics}, vol.~21, no.~4, pp. 2015--2026, Aug 2016.

\bibitem{Hoffmann-Gorinevsky-2007}
G.~M. Hoffmann, D.~Gorinevsky, R.~W. Mah, C.~J. Tomlin, and J.~D. Mitchell,
  ``Fault tolerant relative navigation using inertial and relative sensors,''
  in \emph{Proc. of the AIAA Space 2000 Conference}, Hilton Head, South
  Carolina, USA, 20--27 August 2007.

\bibitem{Aghili-2022}
F.~Aghili, ``Fault-tolerant and adaptive visual servoing for capturing moving
  objects,'' \emph{IEEE/ASME Transactions on Mechatronics}, vol.~27, no.~3, pp.
  1773--1783, 2022.

\bibitem{Kriegsman-Bernard-1966}
B.~A. Kriegsman, ``Radar-updated inertial navigation of a continuously-powered
  space vehicle,'' \emph{Aerospace and Electronic Systems, IEEE Transactions
  on}, vol. AES-2, no.~4, pp. 549--565, July 1966.

\bibitem{Emadzadeh-Speyer-2011}
A.~Emadzadeh and J.~Speyer, ``Relative navigation between two spacecraft using
  x-ray pulsars,'' \emph{Control Systems Technology, IEEE Transactions on},
  vol.~19, no.~5, pp. 1021--1035, Sept 2011.

\bibitem{Aghili-2019e}
F.~Aghili, ``Optimal post-grasping robotic manuvers for satbilization of a
  tumbling stallite,'' \emph{{AIAA} Journal of Guidance, Control, and
  Dynamics}, vol.~43, no.~10, p. 1952?1959, 2020.

\bibitem{Liu-Fang-2015}
J.~Liu, J.~Fang, X.~Ma, Z.~Kang, and J.~Wu, ``X-ray pulsar/starlight doppler
  integrated navigation for formation flight with ephemerides errors,''
  \emph{Aerospace and Electronic Systems Magazine, IEEE}, vol.~30, no.~3, pp.
  30--39, March 2015.

\bibitem{Aghili-2012b}
F.~Aghili, ``Active orbital debris removal using space robotics,'' in
  \emph{International Symposium on Artificial Intelligence, Robotics and
  Automation in Space {i-SAIRAS}}, Turin, Italy, Sep.~4--6 2012.

\bibitem{Aghili-Kuryllo-Okouneva-English-2010a}
F.~Aghili, M.~Kuryllo, G.~Okouneva, and C.~English, ``Fault-tolerant
  position/attitude estimation of free-floating space objects using a laser
  range sensor,'' \emph{IEEE Sensors Journal}, vol.~11, no.~1, pp. 176--185,
  Jan. 2011.

\bibitem{Corazzini-Robertson-1997}
T.~Corazzini, A.~Robertson, A.~Adams, and J.~C. Hassibi, ``Gps sensing for
  spacecraft formation,'' in \emph{Proc. of {ION-GPS}}, Kansas City, MO,
  September 1997, pp. 735--744.

\bibitem{Wolfe-Speyer-2004}
J.~Wolfe and J.~L. Speyer, ``Effective estimation of relative positions in
  orbit using differential carrier-phase,'' in \emph{Proc. of of the AIAA
  Guidance, Navigation and Control Conference}, Providence, Rhode Island,
  August 2004.

\bibitem{Aghili-2013}
F.~Aghili, ``Pre- and post-grasping robot motion planning to capture and
  stabilize a tumbling/driftig free-floater with uncertain dynamics,'' in
  \emph{IEEE International Conf. on Robotics \& Automation}, Karlsruhe,
  Germany, May~6--10 2013, pp. 5441--5448.

\bibitem{Almagbile-Wang-Ding-2010}
A.~Almagbile, J.~Wang, and W.~Ding, ``Evaluating the performance of adaptive
  kalman filter methods in gps/ins integration,'' \emph{Journal of Global
  Positioning Systems}, vol.~9, no.~1, pp. 33--40, 2010.

\bibitem{Masutani-Iwatsu-Miyazaki-1994}
Y.~Masutani, T.~Iwatsu, and F.~Miyazaki, ``Motion estimation of unknown rigid
  body under no external forces and moments,'' in \emph{IEEE Int. Conf. on
  Robotics \& Automation}, San Diego, May 1994, pp. 1066--1072.

\bibitem{Samson-English-Deslauriers-Christie-2004}
C.~Samson, C.~English, A.~Deslauriers, I.~Christie, F.~Blais, and F.~Ferrie,
  ``Neptec 3{D} laser camera system: From space mission {STS}-105 to
  terrestrial applications,'' \emph{Canadian Aeronautics and Space Journal},
  vol.~50, no.~2, pp. 115--123, 2004.

\bibitem{Mokuno-Kawano-2004}
M.~Mokuno, I.~Kawano, and T.~Suzuki, ``In-orbit demonstration of rendezvous
  laser radar for unmanned autonomous rendezvous docking,'' \emph{Aerospace and
  Electronic Systems, IEEE Transactions on}, vol.~40, no.~2, pp. 617--626,
  April 2004.

\bibitem{Aghili-Parsa-2008b}
F.~Aghili and K.~Parsa, ``An adaptive vision system for guidance of a robotic
  manipulator to capture a tumbling satellite with unknown dynamics,'' in
  \emph{{IEEE/RSJ} Int. Conf. on Intelligent Robots and Systems}, Nice, France,
  September 2008, pp. 3064--3071.

\bibitem{Aghili-2008c}
F.~Aghili, ``Optimal control for robotic capturing and passivation of a
  tumbling satellite with unknown dyanmcis,'' in \emph{{AIAA} Guidance,
  Navigation and Control Conference}, Honolulu, Hawaii, August 2008.

\bibitem{Shibata-Matsumoto-1996}
T.~Shibata, Y.~Matsumoto, T.~Kuwahara, M.~Inaba, and H.~Inoue, ``Development
  and integration of generic components for a teachable vision-based mobile
  robot,'' \emph{Mechatronics, IEEE/ASME Transactions on}, vol.~1, no.~3, pp.
  230--236, Sept 1996.

\bibitem{Aghili-2011k}
F.~{Aghili}, ``A prediction and motion-planning scheme for visually guided
  robotic capturing of free-floating tumbling objects with uncertain
  dynamics,'' \emph{IEEE Transactions on Robotics}, vol.~28, no.~3, pp.
  634--649, June 2012.

\bibitem{Lu-Tomizuka-2006}
G.~Lu and M.~Tomizuka, ``Lidar sensing for vehicle lateral guidance: Algorithm
  and experimental study,'' \emph{Mechatronics, IEEE/ASME Transactions on},
  vol.~11, no.~6, pp. 653--660, Dec 2006.

\bibitem{Leavitt-Sideris-2006}
J.~Leavitt, A.~Sideris, and J.~Bobrow, ``High bandwidth tilt measurement using
  low-cost sensors,'' \emph{Mechatronics, IEEE/ASME Transactions on}, vol.~11,
  no.~3, pp. 320--327, June 2006.

\bibitem{Guoqiang-Corradi-2011}
G.~Fu, P.~Corradi, A.~Menciassi, and P.~Dario, ``An integrated triangulation
  laser scanner for obstacle detection of miniature mobile robots in indoor
  environment,'' \emph{Mechatronics, IEEE/ASME Transactions on}, vol.~16,
  no.~4, pp. 778--783, Aug 2011.

\bibitem{Wang-Liu-2014}
K.~Wang, Y.~hui Liu, and L.~Li, ``A simple and parallel algorithm for real-time
  robot localization by fusing monocular vision and odometry/ahrs sensors,''
  \emph{Mechatronics, IEEE/ASME Transactions on}, vol.~19, no.~4, pp.
  1447--1457, Aug 2014.

\bibitem{Simanek-Reinstein-2015}
J.~Simanek, M.~Reinstein, and V.~Kubelka, ``Evaluation of the ekf-based
  estimation architectures for data fusion in mobile robots,''
  \emph{Mechatronics, IEEE/ASME Transactions on}, vol.~20, no.~2, pp. 985--990,
  April 2015.

\bibitem{Aghili-2010f}
F.~Aghili, ``Automated rendezvous \& docking {(AR\&D)} without impact using a
  reliable 3d vision system,'' in \emph{{AIAA} Guidance, Navigation and Control
  Conference}, Toronto, Canada, August 2010.

\bibitem{English-Zhu-Smith-Ruel-Christie-2005}
C.~English, S.~Zhu, C.~Smith, S.~Ruel, and I.~Christie, ``{TriDar}: A huybrid
  sensorfor exploiting the complementary nature of triangulation and lidar
  technologies,'' in \emph{International Symposium on Artificial Intelligence,
  Robotics and Automation in Space ({ISAIRAS})}, Munich, Germany, 5-8 September
  2005.

\bibitem{Aghili-Parsa-Martin-2008b}
F.~Aghili, K.~Parsa, and E.~Martin, ``A vision-guided system for robotic
  capture of a tumbling satellite in presence of occlusion,'' in
  \emph{{CISM-IFToMM} Symposium on Robot Design, Dynamics, and Control}, Tokyo,
  Japan, July 2008, pp. 3--10.

\bibitem{Aghili-Parsa-2007b}
F.~Aghili and K.~Parsa, ``Adaptive motion estimation of a tumbling satellite
  using laser-vision data with unknown noise characteristics,'' in \emph{2007
  IEEE/RSJ International Conference on Intelligent Robots and Systems}, Oct
  2007, pp. 839--846.

\bibitem{Aghili-Parsa-Martin-2008a}
F.~Aghili, K.~Parsa, and E.~Martin, ``Robotic docking of a free-falling space
  object with occluded visual condition,'' in \emph{9th Int. Symp. on
  Artificial Intelligence, Robotics \& Automation in Space}, Los Angeles, CA,
  Feb.~26 -- 29 2008.

\bibitem{Aghili-Kuryllo-Okouneva-English-2010c}
F.~Aghili, M.~Kuryllo, G.~Okouneva, and C.~English, ``Fault-tolerant pose
  estimation of space objects,'' in \emph{IEEE/ASME Int. Conf. on Advanced
  Intelligent Mechatronics (AIM)}, Montreal, Canada, July 2010, pp. 947--954.

\bibitem{Aghili-Kuryllo-Okouneva-English-2010b}
------, ``Robust vision-based pose estimation of moving objects for automated
  rendezvous \& docking,'' in \emph{IEEE Int. Conf. on Mechatronics and
  Automation (ICMA)}, Xian, China, August 2010, pp. 305--311.

\bibitem{Kim-Hwang-2004}
S.-H. Kim, Y.-H. Hwang, H.~K. Hong, and M.-H. Choi, \emph{Advances in
  Artificial Intelligence {MICAI}}.\hskip 1em plus 0.5em minus 0.4em\relax
  Berlin, Heidelberg: Springer, 2004, ch. An Improved {ICP} Algorithm Based on
  the Sensor Projection for Automatic {3D} Registration, pp. 642--651.

\bibitem{Besl-Mckay-1992}
P.~J. Besl and N.~D. McKay, ``A method for registration of {3-D} shapes,''
  \emph{{IEEE} Trans. on Pattern Analysis \& Machine Intelligence}, vol.~14,
  no.~2, pp. 239--256, 1992.

\bibitem{Greenspan-Yurick-2003}
M.~Greenspan and M.~Yurick, ``Approximate {K-D} tree search for efficient
  {ICP},'' in \emph{{IEEE} International Conference on Recent Advances in 3D
  Digital Imaging and Modeling}, Banff, Canada, October 2003, pp. 442--448.

\bibitem{Chen-Medioni-1992}
Y.~Chen and G.~G. Medioni, ``Object modeling by registration of multiple range
  images,'' \emph{Image and Vision Computing}, vol.~10, no.~3, pp. 145--155,
  1992.

\bibitem{Rusinkiewicz-Levoy-2001}
S.~Rusinkiewicz and M.~Levoy, ``Efficient variants of the {ICP} algorithm,'' in
  \emph{Proc. of 3rd Int. Conf. on 3-D Digital Imaging and Modeling}, Quebec
  City, Canada, 28 May -- 1 June 2001, pp. 145--152.

\bibitem{Godin-Laurendeau-Bergevin-2001}
G.~Godin, D.~Laurendeau, and R.~Bergevin, ``A method for the registration of
  attributed range images,'' in \emph{Proc. of 3rd Int. Conf. on 3-D Digital
  Imaging and Modeling}, Quebec City, Canada, 28 May -- 1 June 2001, pp.
  179--186.

\bibitem{Du-Zheng-Ying-Liu-2010}
S.~Du, N.~Zheng, S.~Ying, and J.~Liu, ``Affine iterative closest point
  algorithm for point set registration,'' \emph{Pattern Recognition Letters},
  vol.~31, no.~9, pp. 791--799, 2010.

\bibitem{Dong-Peng-Ying-Hu-2014}
J.~Dong, Y.~Peng, S.~Ying, and Z.~Hu, ``Lietricp: Animprovement of trimmed
  iterative closest point algorithm,'' \emph{Journal of Neurocomputing}, vol.
  140, no.~22, pp. 67--76, September 2014.

\bibitem{Li-Yin-Huang-2011}
W.~Li, Z.~Yin, Y.~Huang, and Y.~Xiong, ``Three-dimensional point-based shape
  registration algorithm based on adaptive distance function,'' \emph{Computer
  Vision, IET}, vol.~5, no.~1, pp. 68--76, Jan 2011.

\bibitem{Pottmann-Hung-Yang-2006}
H.~Pottmann, Q.-X. Hung, Y.-L. Yang, and S.-M. Hu, ``Geometry and convergence
  analysis of algorithms for registration of 3d shapes,'' \emph{Int. Journal of
  Computer Vision}, vol.~67, no.~3, pp. 277--296, 2006.

\bibitem{Wang-Pottmann-Liu-2006}
W.~Wang, H.~Pottmann, and Y.~Liu, ``Fitting b-spline curves to point clouds by
  curvature-based squared distance minimization,'' \emph{{ACM} Trans. Graph.},
  vol.~25, no.~2, pp. 214--238, 2006.

\bibitem{Hillenbrand-Lampariello-2005}
U.~Hillenbrand and R.~Lampariello, ``Motion and parameter estimation of a
  free-floating space object from range data for motion prediction,'' in
  \emph{The 8th Int. Symposium on Artifcial Intelligence, Robotics and
  Automation in Space: i-SAIRAS 2005}, Munich, Germany, Sep.~5--8 2005.

\bibitem{Clohessy-Wiltshire-1960}
W.~H. Clohessy and R.~S. Wiltshire, ``Terminal guidance system for satellite
  rendezvous,'' \emph{Journal of Aerosapce Science}, vol.~27, no.~9, pp.
  653--658, 1960.

\bibitem{Lefferts-Markley-Shuster-1982}
E.~J. Lefferts, F.~L. Markley, and M.~D. Shuster, ``Kalman filtering for
  spacecraft attitude estimation,'' vol.~5, no.~5, pp. 417--429, Sep.--Oct.
  1982.

\bibitem{Pittelkau-2001}
M.~E. Pittelkau, ``Kalman filtering for spacecraft system alignment
  calibration,'' vol.~24, no.~6, pp. 1187--1195, Nov. 2001.

\bibitem{Wilcox-1967}
J.~C. Wilcox, ``A new algorithm for strapped-down inertial navigation,''
  \emph{IEEE Trans. on Aerospace and Electronic Systems}, vol.~3, no.~5, pp.
  796--802, Sep. 1967.

\bibitem{Zanetti-2012}
R.~Zanetti, ``Recursive update filtering for nonlinear estimation,''
  \emph{Automatic Control, IEEE Transactions on}, vol.~57, no.~6, pp.
  1481--1490, June 2012.

\bibitem{Valappil-Georgakis-2000}
J.~Valappil and C.~Georgakis, ``Systematic estimation of state noise statistics
  for extended kalman filters,'' \emph{AIChE Journal}, vol.~46, no.~2, pp.
  292--308, Feb 2000.

\bibitem{vanLoan-1978}
C.~F. van Loan, ``Computing integrals involving the matrix exponential,''
  \emph{{IEEE} Trans. on Automatic Control}, vol.~23, no.~3, pp. 396--404, Jun.
  1978.

\bibitem{Chui-Chen-1998-p113}
C.~K. Chui and G.~Chen, \emph{Kalman Filtering with Real-Time
  Applications}.\hskip 1em plus 0.5em minus 0.4em\relax Berlin: Springer, 1998,
  pp. 113--115.

\bibitem{Mehra-1970}
R.~Mehra, ``On the identification of variances and adaptive kalman filtering,''
  \emph{Automatic Control, IEEE Transactions on}, vol.~15, no.~2, pp. 175--184,
  Apr 1970.

\bibitem{Wang-2000}
J.~Wang, ``Stochastic modeling for real-time kinematic gps/glonass position,''
  \emph{Journal of Navigation}, vol.~46, no.~4, pp. 297--305, 2000.

\bibitem{Kim-Jee-Park-Lee-2009}
K.~Kim, G.~Jee, C.-G. Park, and J.-G. Lee, ``The stability analysis of the
  adaptive fading extended kalman filter using the innovation covariance,''
  \emph{International Journal of Control, Automation and Systems}, vol.~7,
  no.~1, pp. 49--56, Feb 2009.

\bibitem{Gao-Wei-Zhong-Subic-2015}
S.~Gao, W.~Wei, Y.~Zhong, and A.~Subic, ``Sage windowing and random weighting
  adaptive filtering method for kinematic model error,'' \emph{Aerospace and
  Electronic Systems, IEEE Transactions on}, vol.~51, no.~2, pp. 1488--1500,
  April 2015.

\bibitem{Aghili-Parsa-2009}
F.~Aghili and K.~Parsa, ``Motion and parameter estimation of space objects
  using laser-vision data,'' \emph{{AIAA} Journal of Guidance, Control, and
  Dynamics}, vol.~32, no.~2, pp. 538--550, March 2009.

\bibitem{Shang-Jasiobedzki-2005}
L.~Shang, P.~Jasiobedzki, and M.~Greenspan, ``Discrete pose space estimation to
  improve icp-based tracking,'' in \emph{3-D Digital Imaging and Modeling,
  2005. 3DIM 2005. Fifth International Conference on}, June 2005, pp. 523--530.

\bibitem{Simon-Herbert-Kanade-1994}
D.~A. Simon, M.~Herbert, and T.~Kanade, ``Real-time 3-d estimation using a
  high-speed range sensor,'' in \emph{{IEEE} Int. Conference on Robotics \&
  Automation}, San Diego, CA, May 1994, pp. 2235--2241.

\bibitem{Faugeras-Herbert-1986}
O.~D. Faugeras and M.~Herbert, ``The representation, recognition, and locating
  of 3-d objects,'' \emph{The International Journal of Robotics Research},
  vol.~5, no.~3, pp. 27--52, 1986.

\bibitem{Eggert-Lorusso-Fisher-1997}
D.~Eggert, A.~Lorusso, and R.~B. Fisher, ``Estimating {3-D} rigid body
  transformation: a comparison of four major algorithms,'' \emph{Machine Vision
  \& Apllications}, vol.~9, no.~5, March 1997.

\bibitem{Horn-1987}
B.~K.~P. Horn, ``Closed-form solution of absolute orientation using unit
  quaternions,'' \emph{J. Opt. Soc. Amer.}, vol.~4, no.~4, pp. 629--642, Apr.
  1987.

\bibitem{Li-Yin-Huang-Xiong-2011}
Y.~Z. H.~Y. Li, W. and Y.~Xiong, ``Automatic registration for 3d shapes using
  hybrid dimensionality-reduction shape descriptions,'' \emph{Pattern
  Recognition}, vol.~44, no.~12, pp. 2926--2943, December 2011.

\bibitem{Kim-Lee-Cho-Kim-2011}
C.~Kim, J.~Lee, M.~Cho, and C.~Kim, ``Fully automated registration of 3d cad
  model with point cloud from construction site,'' in \emph{Proceedings of the
  28th International Symposium on Automation and Robotics in Construction},
  Seoul, Korea, 29 June--2 July 2011, pp. 917--922.


\end{thebibliography}


\end{document}